\documentclass[a4paper,aps,prb,twocolumn,amsmath,superscriptaddress]{revtex4-2}
\usepackage{graphicx}
\usepackage{epstopdf}
\usepackage{xcolor}

\begin{document}

\title{Optical probing of the carriers-mediated coupling of the spin of two Co atoms in a quantum dot}

\author{L.~Besombes}\email{lucien.besombes@neel.cnrs.fr}
\affiliation{Institut Néel, CNRS, Univ. Grenoble Alpes and Grenoble INP, 38000 Grenoble, France}
\author{J.~Kobak}
\affiliation{Univ. of Warsaw, Faculty of Physics, ul. Pasteura 5, 02-093 Warsaw, 
Poland}
\author{W.~Pacuski}
\affiliation{Univ. of Warsaw, Faculty of Physics, ul. Pasteura 5, 02-093 Warsaw, 
Poland}

\date{\today}

\begin{abstract}

We report on the optical spectroscopy of the spin of two Co atoms in a quantum dot and interacting with a single exciton. The spectrum of quantum dots containing two Co atoms are exchange interaction and by the strain at the location of the magnetic atoms. A wide range of spectrum can be obtained depending on the relative coupling of each atom to the confined exciton. We obtained a comprehensive interpretation of the experimental data with a spin Hamiltonian model. We show that the two Co atoms spins can be orientated by the injection of spin polarized carriers at zero magnetic field. This induces a correlation between the two spins that is observed in the intensity distribution of the emission spectra. The optical absorption in the phonon sideband of quantum dots doped with two Co reveals resonant absorptions which strongly depend on a transverse magnetic field. We show that these characteristic absorptions result from an interplay between the mixing of Co spin states induced by the presence of in-plane strain anisotropy at the magnetic atoms location and the transverse field. 

\end{abstract}

\maketitle

\section{Introduction}

Individual localized spins in semiconductors, which can feature both electrical and optical control, hold promise for solid-state quantum information storage. Optically active semiconductor quantum dots (QDs) containing an individual carrier offer the possibility of an efficient spin-photon interface, an essential element towards the establishment of quantum information networks \cite{Lukin2019,Senell2022}. Localized spins of transition metal elements can also be addressed optically when they are inserted in a QD \cite{Leger2005, Kudelski2007, Bhattacharjee2007, Kobak2014, Besombes2012}. This spin-photon interface is established by the exchange interaction of magnetic atoms with the spin of optically created carriers. A wide range of magnetic atoms can be inserted into semiconductors, offering optical access to a wide choice of localized electron spin, nuclear spin and orbital momentum \cite{Kobak2014}.

Semiconductor heterostructures offer in addition the unique possibility of controlling the exchange interaction between distant magnetic dopants by varying the carrier density of the host \cite{Dietl2000}. In the case of a dot containing two or a few magnetic atoms, the spins are only coupled via a short range exchange that could be only relevant for the first or second neighbors. The injection of a single exciton whose coupling with the magnetic atoms is dominated by the exchange interaction with the hole spin will couples ferromagnetically the atoms. This has been demonstrated in the case of two manganese atoms (Mn) in both II-VI and III-V semiconductors \cite{Besombes2012,Krebs2013}.  

We analyze here the case of cobalt (Co) in II-VI semiconductors, a magnetic element which carries a localized spin with a large sensitivity to local strain. Co, a 3$d^7$ transition element, is usually incorporated in II-VI semiconductors as a Co$^{2+}$ ion carrying a spin S=3/2 with an orbital momentum L=3. All the stable isotopes also carries a nuclear spin I=7/2. The orbital momentum efficiently connects the Co spin to its strain environment through the modulation of the crystal field and the spin-orbit coupling making Co an interesting spin qubit for hybrid spin-mechanical systems \cite{Fuch2015,Rabl2010,Ovart2014,Schuetz2015,Besombes2019}. 

The influence of static local strain on the emission of QDs doped with a single Co atom has been analyzed \cite{Kobak2014}. Here we demonstrate that in this diluted magnetic semiconductor system the exchange coupling is high enough to permit to spectrally resolve the different magnetic ground states of two Co atoms and address them independently. We performed the magneto-optic spectroscopy and resonant optical spectroscopy of CdTe/ZnTe QDs doped with two Co atoms. This permits first to clearly identify the main parameters controlling the characteristic emission spectrum of such QDs. We also show that the two Co spins can be optically oriented by the injection of spin polarized carriers. The induced correlation between the localized spins is observed in the photoluminescence (PL) intensity distribution. Under resonant optical excitation, the absorption of the QDs' phonon sideband reveals transitions which strongly depend on a transverse magnetic field. We show that these characteristic absorptions result from an interplay of the spin states mixing induced by the presence of an in-plane strain anisotropy at the magnetic atoms location and the transverse magnetic field. 

The rest of the paper is organized as follows: After a short presentation of the sample and experiments in Sec. II, we discuss in Sec. III the magneto-optic properties of QDs containing two Co atoms. The fine structure of a confined exciton in the exchange field of two Co atoms is analyzed and modeled in detail. In Sec. IV we show that the two Co spins are correlated by there mutual interaction with confined carriers. We demonstrate that the two Co spins can be oriented by the optical injection of spin polarized carriers and measured spin fluctuations under optical excitation in the tens of nanosecond range. In Sec. V we present and discuss the transverse magnetic field dependence of the acoustic phonon sideband absorption. The conclusion is drawn in Sec. VI.

\section{Experimental details}

In this study we use self-assembled CdTe/ZnTe QDs grown by molecular beam epitaxy and doped with Co atoms. The concentration of Co was adjusted to obtain QDs containing 0, 1, 2 or a few magnetic atoms \cite{Kobak2014}.

Individual QDs are studied by optical micro-spectroscopy at liquid helium temperature (T= 4.2 K). The sample is mounted on x,y,z piezo actuators and inserted in a vacuum tube under a low pressure of He exchange gaz. The tube is immersed in the variable temperature insert of the cryostat filled with liquid He. The sample temperature is measured thanks to a sensor located in the copper sample holder. The cryostat is equipped with a vectorial superconducting coil and a magnetic field up to 9 T along the QD growth axis and 2T in the QD plane can be applied.

The PL is excited with a tunable dye laser and collected by a high numerical aperture microscope objective (NA=0.85). For spectral analysis, the PL is dispersed by a two meters double grating spectrometer (1800 gr/mm) and detected by a cooled Si charge-coupled-device camera. Power dependence measurements are performed by using a neutral gradient filter. For the Photo-Luminescence Excitation (PLE) measurements, the power of the tunable dye laser is stabilized with an electro-optic variable attenuator when its wavelength is tuned.

A half wave plate in front of a linear polarizer is used to analyze the linear polarization. For circular polarization measurement, a quarter-wave plate oriented at 45° from the linear polarization direction of detection is inserted in the detection path. When inserted both in the excitation and detection paths, at 45° from the linear polarization of the excitation laser, the quarter-wave plate, combined with a half-wave plate and a linear polarizer, is used to select co-circularly or cross-circularly polarized PL. 

For the auto-correlation measurements, a single QD emission line is selected by the exit slits of the double spectrometer and the PL signal sent to an Hanbury Brown–Twiss (HBT) setup. The HBT setup is equipped with two Si avalanche photodiodes in combination with a time-correlated photon-counting module giving an overall time resolution of around 700 ps.

\section{Individual QDs doped with two Co atoms}

In the studied sample, a large fraction of the dots have no magnetic atom inside, some of them have 1 Co and in some instance 2 Co or more. Dots with a single Co sufficiently coupled to an exciton give rise to PL with 4 peaks. The intensity distribution within these four peaks changes from dot to dot with in general a weaker intensity or even no PL detectable on the inner two lines \cite{Kobak2014}.

The origin of the 4 peaks can be accounted for by the strongly anisotropic coupling of the exciton to the four electronic spin states of the Co $S_z$=$\pm1/2$,$\pm3/2$. The Co-exciton interaction is dominated by the coupling with the spin of the hole. In theses self-assembled QDs, the holes have a heavy-hole character with a spin J$_{hz}$=$\pm3/2$ along the growth axis. This results in an effective magnetic field on the exciton oriented along the QDs' growth direction and given by $g\mu_{B}B_{eff}$=I$_{hCo}$S$_z$ where I$_{hCo}$ is the hole-Co exchange. I$_{hCo}\propto\beta\vert\Psi_h(\vec{r}_{Co})\vert^2$ depends both on the value of the hole wave function at the location $\vec{r}_{Co}$ of the Co in the QD and on $\beta$ a material dependent quantity  \cite{Furdyna1988}. For Co in CdTe $\beta$ is dominated by the kinetic exchange and is antiferromagnetic \cite{Kacman2001}.

\begin{figure*}[hbt]
    \centering
    \includegraphics[width=0.9\linewidth]{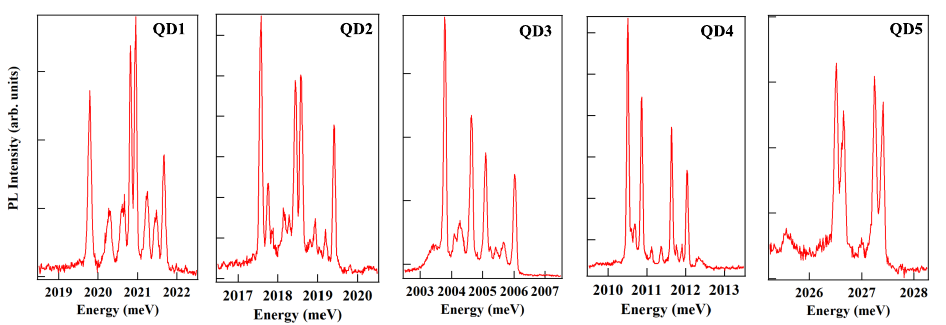} 
    \caption{Low temperature (T=4.2K) photoluminescence spectra of five QDs containing two Co atoms labeled from QD1 to QD5. The diversity of spectra is controlled by the ratio of exchange couplings of the two atoms with the exciton and by the strained induced splitting of the spin of each atom.} 
    \label{fig:PL} 
\end{figure*}

The intensity distribution observed in Co-doped QDs arises from the strain induced fine structure of the magnetic atom. The strain influences the fine structure of the atom via the deformation of the crystal field and the spin-orbit coupling. The two inner lines of the exciton-Co system are associated with the Co spin states S$_z$=$\pm1/2$. In self-assembled QDs, the strain at the magnetic atom position changes from dot to dot \cite{Kobak2018} but is usually dominated by negative in-plane bi-axial strain. In this case, the S$_z$=$\pm1/2$ spin states are shifted to a higher energy and their occupancy probability at cryogenic temperature is weak.

\subsection{Photoluminescence of QDs containing two Co atoms}

For a dot with two Co atoms, similar qualitative arguments suggest that the dominant effective exchange field felt by the exciton is now $g\mu_{B}B_{eff}$=I$_{hCo_{(1)}}$S$_{z(1)}$+I$_{hCo_{(2)}}$S$_{z(2)}$ which can take up to 16 values. The observation of dots with two Co strongly coupled to the exciton should then give rise to more than 4 but less than 16 lines. 

Emission of different Co-doped QDs measured at T=4.2 K and at zero magnetic field are presented in Fig.~\ref{fig:PL}. The PL presents in general four dominant lines with similar intensities. The splitting between the four main lines is irregular and also  changes from dot to dot. The two inner lines can almost overlap at the center of the structure, as in QD1, or be very close to the two outer lines, forming two doublets, as in QD5. This contrasts with dots doped with a single Co where the splitting is regular and the intensity of the inner and outer lines usually very different \cite{Kobak2018}. 

\begin{figure}[hbt]
    \centering
    \includegraphics[width=1.0\linewidth]{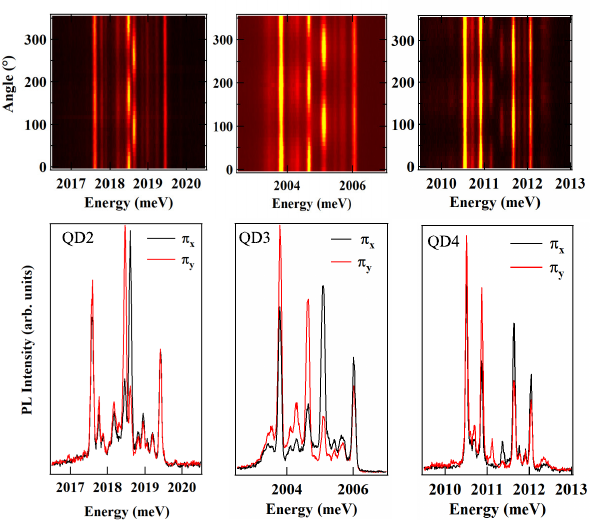} 
    \caption{Top: Linear polarization PL intensity map of QD2, QD3 and QD4. Bottom: Corresponding linearly polarized PL spectra recorded along two orthogonal directions. The direction of polarization is measured with respect to the [100] or [010] axis of the sample (i.e., at 45° from the easy cleavage axis of the substrate).} 
    \label{fig:PLPi} 
\end{figure}

As can be seen from the PL intensity maps of linear polarization in Fig.~\ref{fig:PLPi}, the central lines are significantly linearly polarized and their polarization rate is larger in dots where their energy spacing is smaller. The linear polarization rate is also generally much lower for the two outer lines. 

\begin{figure}[hbt]
    \centering
    \includegraphics[width=1.0\linewidth]{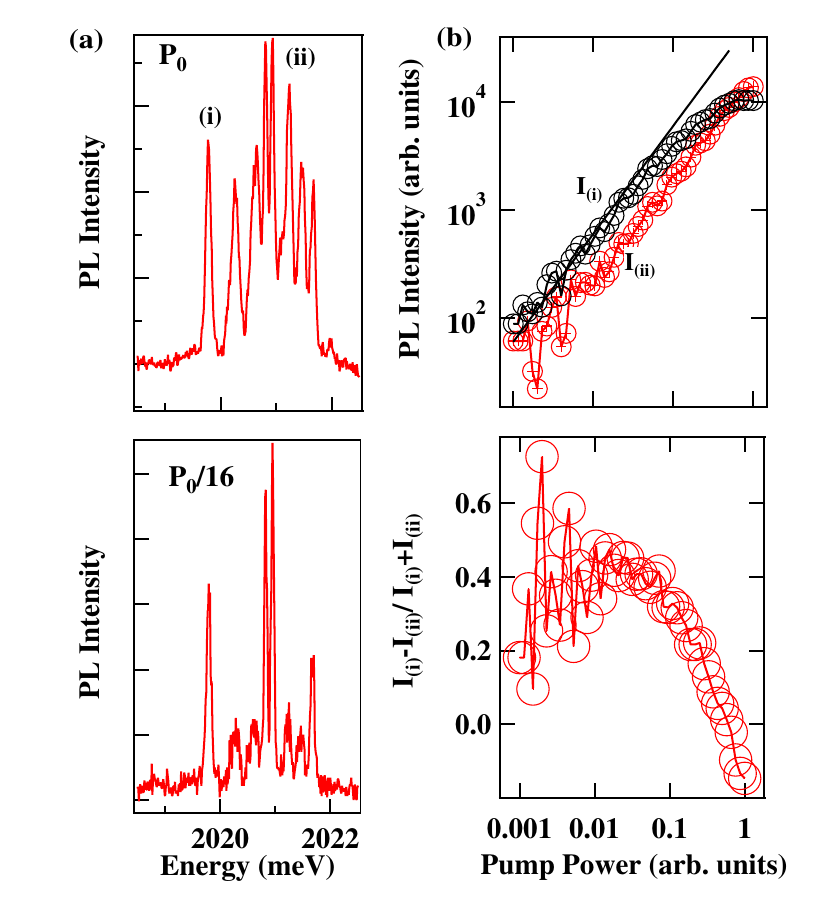} 
    \caption{Excitation power dependence of the PL intensity distribution in QD1 at zero magnetic field. (a) PL spectra for two different excitation powers. (b) Top: Excitation power dependence of the integrated PL intensity of the lines (i) and (ii). Bottom: Corresponding normalized PL intensity ratio (I$_{(i)}$-I$_{(ii)}$)/(I$_{(i)}$+I$_{(ii)}$).} 
    \label{fig:PW} 
\end{figure}

Weaker intensity optical transitions are in general observed between the main PL lines. They can also present some linear polarization. The intensity of these weaker transitions depends on the excitation power. As it can be seen in Fig.~\ref{fig:PW}, their observations is easier at high excitation power and their contributions almost disappear at very low excitation. Such weaker intensity lines between the four main peaks are not observed in QDs containing a single Co atom. 

In the large diversity of observed PL spectrum (see Fig.~\ref{fig:PL}), certain common feature appear: four dominant PL lines and lower intensity lines with a relative contribution that depends on the excitation power. This PL structure is characteristic of dots containing two Co atoms under biaxial in-plane strain. In the presence of negative biaxial in-plane strain, the spins of each Co thermalizes on their ground states S$_{z(i)}$=$\pm$3/2. There are then four equivalent spin ground states $\vert\pm\frac{3}{2}\rangle_{(1)}\vert\pm\frac{3}{2}\rangle_{(2)}$ with total spin M$_z$=$\pm$3 and $\vert\pm\frac{3}{2}\rangle_{(1)}\vert\mp\frac{3}{2}\rangle_{(2)}$ with M$_z$=0. These states are degenerate at zero magnetic field and have the same occupation probability. The other 12 two Co spins configurations contain at least a component S$_{z(i)}$=$\pm$1/2 which is shifted towards high energy by local biaxial strain. The probability of occupancy of these 12 excited states at cryogenic temperature is lower than for the 4 ground states.

\begin{figure}[hbt]
    \centering
    \includegraphics[width=1.0\linewidth]{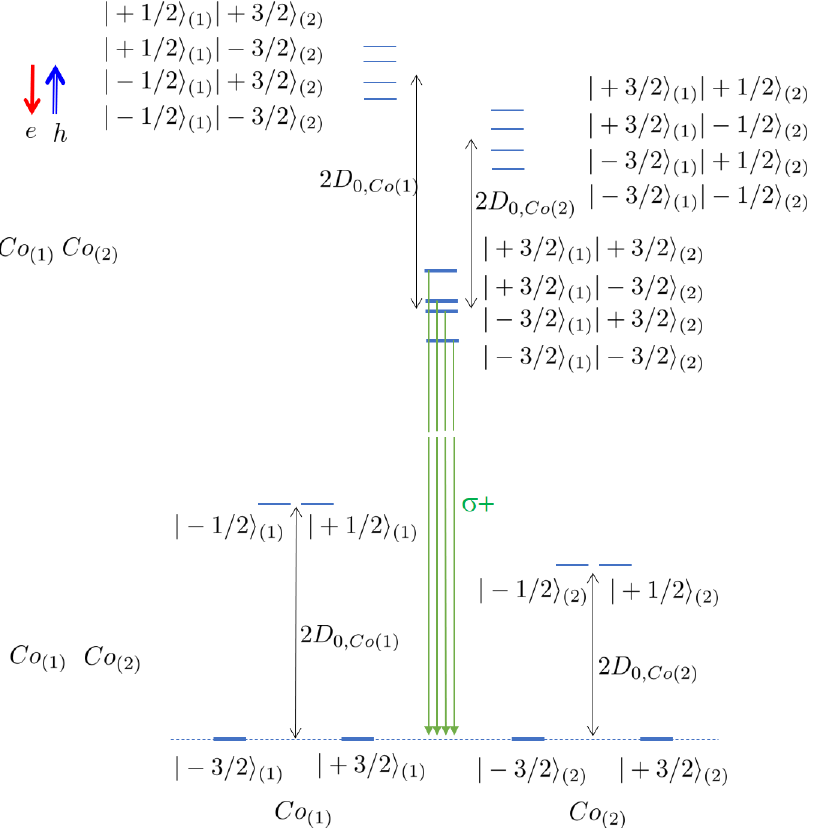} 
    \caption{(a) Scheme of the energy levels at zero magnetic field of a QD containing, in the ground state, two Co atoms with D$_{0,Co(1)}$ $>$ D$_{0,Co(2)}$ and, in the excited state, a $\sigma+$ exciton with I$_{hCo(1)}$ $>$ I$_{hCo(2)}$. Only the PL transitions from the four lowest energy states with the largest occupation probability are presented.} 
    \label{fig:scheme} 
\end{figure}

When interacting with an exciton, the four 2Co ground states which are initially degenerate, are split by the exchange interaction with the confined carriers (see the energy level scheme in Fig.~\ref{fig:scheme}). The resulting four exciton-2Co levels give the main contribution in the PL spectra and four lines with similar intensities are expected. The lowest and the highest energy lines corresponds to, respectively, an anti-ferromagnetic and a ferromagnetic coupling of the hole with the 2Co states $\vert\pm\frac{3}{2}\rangle_{(1)}\vert\pm\frac{3}{2}\rangle_{(2)}$. The two inner lines correspond to the coupling of the exciton with the 2Co states $\vert\pm\frac{3}{2}\rangle_{(1)}\vert\mp\frac{3}{2}\rangle_{(2)}\rangle$. These states with total spin M$_z$=0 generally interact with the exciton as I$_{hCo_{(1)}}$ and I$_{hCo_{(2)}}$, the exchange interaction of the hole with each Co, are different resulting in a non-zero exchange coupling of the hole with the two atoms.

If one neglects the exchange interaction with the electron, I$_{eCo(i)}$, usually much weaker, the relative position of the inner and outer lines depends on the difference I$_{hCo(1)}$-I$_{hCo(2)}$. For example, with I$_{hCo_{(1)}}$=I$_{hCo_{(2)}}$, the exchange interaction of the exciton with the 2 atoms in the configuration M$_z$=0 cancels out and a single line would appear in the middle of the structure. As these lines come closer when I$_{hCo(1)}$-I$_{hCo(2)}$ decreases, any weak anisotropic coupling term that can mix $\sigma+$ and $\sigma-$ excitons becomes significant and induces a linear polarization rate (see QD2, QD3 in Fig.~\ref{fig:PLPi}). On the contrary, for very different values of I$_{hCo_{(1)}}$ and I$_{hCo_{(2)}}$, the dominant inner and outer lines come closer together and can even merge when one of the exchange term tends towards zero. This gives rise to spectrum dominated by two doublets separated by a large central gap (see QD5 in Fig.~\ref{fig:PL}).  

With these simple considerations, in a QD containing two Co atoms, the splitting of the two outer lines is given by 9/2(I$_{hCo(1)}$+I$_{hCo(2)}$) and the splitting of the two inner lines by 9/2(I$_{hCo(1)}$-I$_{hCo(2)}$). The splitting of the inner lines is also likely to be significantly affected by the exchange interaction and possible valence band mixing that can couple $\sigma+$ and $\sigma-$ excitons.

The other 12 2Co states also interact with the exciton. As they contain at least one spin component S$_{z(i)}$=$\pm1/2$, their splitting induced by the exchange interaction with the exciton are different than for the 4 ground states (see Fig.~\ref{fig:scheme}). The occupation probability of these sates is weaker at low temperature and they give rise to lower intensity lines. These PL lines lie between the transition towards the low-energy anti-ferromagnetic states and the transition towards the high-energy ferromagnetic states. 

The contribution in the PL spectra of these higher energy states is expected to increase with the increase of the effective spin temperature of the two Co atoms. An increase of the spin temperature can be induced by the injection of high energy carriers \cite{Besombes2008}. For Co which presents a large spin to strain coupling, it can come from an interaction with non-equilibrium phonons generated during the carriers' relaxation \cite{Tiwari2020}. Such modification of the effective spin temperature is consistent with the increase of the number of observable lines at high excitation power (see Fig.\ref{fig:PW} in the case of QD1). Among these weaker intensity lines, the largest contribution in the PL comes from exciton associated with states containing at least one component $S_{z(i)}=\pm3/2$ of one of the atoms Co$_{(i)}$. These states are the first to be significantly occupied when the spin temperature increases. If both atoms experience large bi-axial strain, only the four ground states contribution will be observed at low temperature.

In 2Co-doped QDs, a large diversity of spectrum can then be expected depending on the position of the two atoms inside the dot and on the strain state at each magnetic atom location. In addition, as observed in other magnetic systems, weak intensity lines can also appear due to the recombination of dark excitons. The presence of anisotropy in the hole-Co exchange interaction and local deviation from a pure bi-axial strain around the atom could allow a recombination of the exciton with a change of the spin state of the magnetic atom and also increase the number of observed lines.

\subsection{Magneto-optics of the exciton coupled with two Co atoms}

\begin{figure}[hbt]
    \centering
    \includegraphics[width=1.0\linewidth]{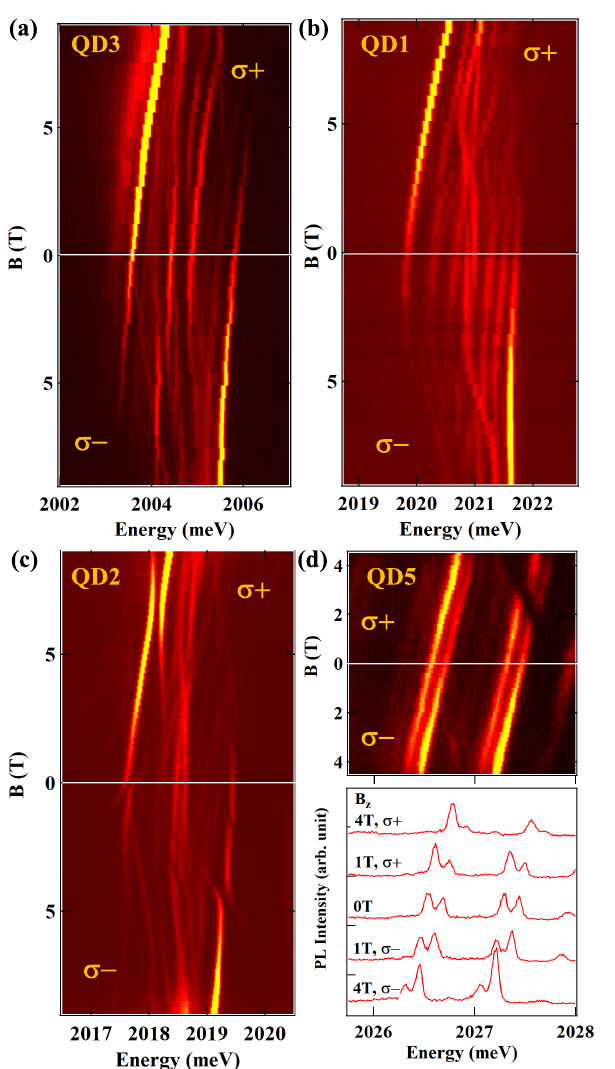} 
    \caption{Intensity map of the longitudinal magnetic field dependence of the PL of four QDs containing two Co atoms, (a) QD3, (b) QD1, (c) QD2 and (d) QD5. In (d), low magnetic field spectra showing the evolution of the intensity distribution among the two doublets of QD5 are presented in the bottom panel.} 
    \label{fig:MagnetoPL1} 
\end{figure}

Magneto-optics measurements permit to extract information on the carrier / magnetic atoms exchange interaction. The longitudinal magnetic field dependence of the PL of four QDs containing two Co atoms are presented in Fig.~\ref{fig:MagnetoPL1} in two circular polarizations. As for the zero field spectra, a large diversity in the magnetic field dependence is observed but some common feature can be extracted. 

The four main lines present a Zeeman splitting and a slight diamagnetic shift. The weaker intensity lines, although sometime presenting anti-crossings, follow a similar longitudinal magnetic field dependence with comparable Zeeman splitting and energy shifts. 

\begin{figure}[hbt]
    \centering
    \includegraphics[width=1.0\linewidth]{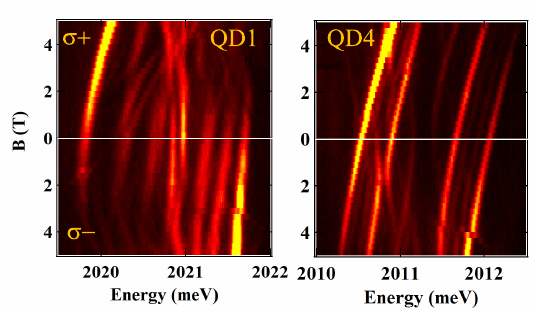} 
\caption{Detail at low magnetic field of the intensity map of the emission of two QDs having a very different ratio I$_{hCo(1)}$/I$_{hCo(2)}$: QD1 with I$_{hCo(1)}$ $\approx$ I$_{hCo(2)}$ and QD4 with I$_{hCo(1)}$ $\gg$ I$_{hCo(2)}$.} 
    \label{fig:MagnetoPL2} 
\end{figure}

We can also notice in the PL intensity maps of Fig.~\ref{fig:MagnetoPL1} that the intensity distribution among the lines is very sensitive to the applied magnetic field. Whereas the four main lines are clearly observed at zero field, at large field the PL mainly arises from one line in each circular polarization. A maximum of PL is observed on the high energy line in $\sigma$- polarization and on the low energy line in $\sigma$+ polarization. This intensity distribution reflects the efficient thermalisation on the 2Co spins states split by the Zeeman energy. With a Lande factor of the Co $g_{Co}\approx2.5$ \cite{Kobak2018}, the two Co spins thermalizes on the states S$_{z(i)}$=-3/2. For an anti-ferromagnetic coupling with the exciton, this corresponds to a low energy $\sigma$+ exciton state $\vert\downarrow_e,\Uparrow_h\rangle\vert-\frac{3}{2}\rangle_{(1)}\vert-\frac{3}{2}\rangle_{(2)}$ and to a high energy $\sigma$- exciton state $\vert\uparrow_e,\Downarrow_h\rangle\vert-\frac{3}{2}\rangle_{(1)}\vert-\frac{3}{2}\rangle_{(2)}$, as observed in the experiment under large magnetic field.

A change in the intensity distribution is also observed for the two inner lines. In dots with very different I$_{hCo(i)}$ under magnetic field ({\it i.e.} PL spectrum with two doublets separated by a central gap, Fig.~\ref{fig:MagnetoPL1}(d)), most of the PL comes from the high energy line in $\sigma-$ and low energy line in $\sigma+$ but a change in the intensity distribution is also observed among the doublets. The high energy line of the doublet dominates in $\sigma-$ polarization whereas most of the intensity comes from the low energy line of the doublet in $\sigma+$ polarization (Fig.~\ref{fig:MagnetoPL1}(d)).

To explain this behavior, let's first consider that I$_{hCo(1)}\gg$ I$_{hCo(2)}$. This corresponds to the case of QD5 or QD4. For a $\sigma-$ exciton, the high energy line is associated with $\vert-\frac{3}{2}\rangle_{(1)}\vert-\frac{3}{2}\rangle_{(2)}$. The second line of the high energy doublet corresponds to $\vert-\frac{3}{2}\rangle_{(1)}\vert+\frac{3}{2}\rangle_{(2)}$. Under a positive magnetic field, the occupation probability of the state S$_{z(2)}$=+3/2 decreases and consequently the PL intensity of the line associated with $\vert-\frac{3}{2}\rangle_{(1)}\vert+\frac{3}{2}\rangle_{(2)}$ becomes weaker than the high energy line of the doublet. In the opposite circular polarization, $\sigma+$, the high energy state is now associated with $\vert+\frac{3}{2}\rangle_{(1)}\vert+\frac{3}{2}\rangle_{(2)}$. Under magnetic field its occupation probability becomes weaker than for $\vert+\frac{3}{2}\rangle_{(1)}\vert-\frac{3}{2}\rangle_{(2)}$ and the intensity distribution among the high energy doublet is reversed, as observed in the experiment (Fig.~\ref{fig:MagnetoPL1}(d)).

When I$_{hCo(1)}$ $\approx$ I$_{hCo(2)}$, the two dominant inner lines are close to the center of the spectrum. However, the magnetic field dependence of their intensity distribution is similar. This gives rise to a configuration in which the distribution of intensity between the two central lines is opposite to that of the two outer lines (see Fig.~\ref{fig:MagnetoPL2}).

\subsection{Model of the magneto-optic properties of QDs containing two Co atoms}

Many parameters are necessary to describe the details of the interaction of confined carriers with two Co spins in a QD. In addition to the confined carriers spins fine structure, the position of each atom and the detail of the strain distribution at the magnetic atom's location can influence the spectrum. It is hardly possible to independently adjust all these parameters and we chose here to discuss a simplified model which can explain the main feature of the observed spectrum. 

\begin{table*}[!hbt]
    \center
\begin{tabular}{c|cccccccccccc}
            & I$_{hCo(i)}$ & I$_{eCo(i)}$ & D$_{0(i)}$ & E$_{(i)}$ &  g$_{Co(i)}$ & \\
            & $\mu$eV & $\mu$eV  & meV  & meV &  \\
            \hline \hline
Co$_{(1)}$  & 290  & 0 & -1.5 & -0.2 & 2.5 &   \\
Co$_{(2)}$  & 210  & 0 & -1.5 & -0.2 & 2.5 &    \\
            \hline \hline
            & 2/3$\delta_{0}$  & $\Delta_{lh}$ & $\rho_s$ & $\theta_s$ & $\delta_1$ & $\gamma$          & g$_e$  & g$_h$  & $\delta_{xz}$ & $\delta_{yz}$ \\
            & $\mu$eV   &        meV    &     meV  &  $^\circ$ & $\mu$eV    & $\mu$eV$T^{-2}$   &   & &   meV  &    meV      \\
            \hline \hline
    X       & -650 & 25 &   1 &  0 &  100  &   2   & -0.2   &  0.4 & 0 & 0   \\
            \hline \hline
\end{tabular}
    \caption{Parameters used in the modeling presented in figure \ref{fig:model0B} and \ref{fig:modelB}. The chosen exchange parameters reproduces the overall splitting of QD3. The other parameters are typical for individual Co \cite{Kobak2018} or individual CdTe/ZnTe QDs \cite{Lafuente2016}.}
    \label{TablePar}
\end{table*}

The electron-hole-2Co Hamiltonian in a self-assembled QD, $\mathcal{H}_{X,2Co}$, can be separated into five parts: 

\begin{eqnarray}
\mathcal{H}_{X,2Co}=\sum_i(\mathcal{H}_{s,Co(i)}+\mathcal{H}_{c,Co(i)})+\mathcal{H}_{B}+\mathcal{H}_{eh}+\mathcal{H}_{bd}
\label{Hamilton}
\end{eqnarray}

$\mathcal{H}_{s,Co(i)}$ describes the fine structure of the Co atoms and their dependence on local strain. For dominant bi-axial in-plane strain, the induced zero field splitting of the S=3/2 states specific to each Co$_{(i)}$ (i=1,2) reads:

\begin{eqnarray}
\mathcal{H}_{s,Co(i)}=\frac{2}{3}D_{0(i)}\left[S_{z_i}^2-\frac{1}{2}\left(S_{x_i}^2+S_{y_i}^2\right)\right]+\frac{E_i}{2}\left(S_{x_i}^2-S_{y_i}^2\right) 
\end{eqnarray}

\noindent where D$_{0(i)}$ splits S$_{z(i)}$=$\pm$3/2 and S$_{z(i)}$=$\pm$1/2 spin states and E$_{(i)}$ mix spin states S$_{z(i)}$=$\pm$3/2 and S$_{z(i)}$=$\mp$1/2 respectively.

The second term in (\ref{Hamilton}), $\mathcal{H}_{c,Co(i)}$, describes the exchange coupling of the electron and hole spins with the Co spins. In this model, a Heisenberg-type exchange interaction is assumed between both Co spins and the hole and electron spins:

\begin{eqnarray}
\mathcal{H}_{c,Co(i)}=I_{hCo(i)}\vec{S_i}.\vec{J_h}+I_{eCo(i)}\vec{S_i}.\vec{\sigma_e} 
\end{eqnarray}

\noindent with $I_{hCo(i)}$ and $I_{eCo(i)}$ the exchange integrals of the hole ($\vec{J_h}$) and the electron ($\vec{\sigma_e}$) spins with the Co spins ($\vec{S_i}$). 

A magnetic field couples via the Zeeman terms to both Co spins and carriers spins and a diamagnetic shift of the electron-hole pair can also be included resulting in

\begin{eqnarray}
\mathcal{H}_{B}=\sum_ig_{Co(i)}\mu_B\vec{B}.\vec{S_i}+g_e\mu_B\vec{B}.\vec{\sigma_e}+g_h\mu_B\vec{B}.\vec{J_h}+\gamma B^2
\end{eqnarray}

\noindent where g$_e$, g$_h$ and g$_{Co(i)}$ are the electron, hole and Co Lande factors and $\gamma$ a diamagnetic coefficient.

The electron-hole exchange interaction, $\mathcal{H}_{eh}$, contains both short-range and long-range parts. We consider here QDs with a C$_{2v}$ symmetry (ellipsoidal flat lenses for instance) where the short-range and the long-range parts contribute to a splitting $\delta_0$ of the bright and dark excitons and where the long-range part induces a coupling $\delta_1$ between the bright excitons \cite{Lafuente2016}.

\begin{figure}[hbt]
    \centering
    \includegraphics[width=1.0\linewidth]{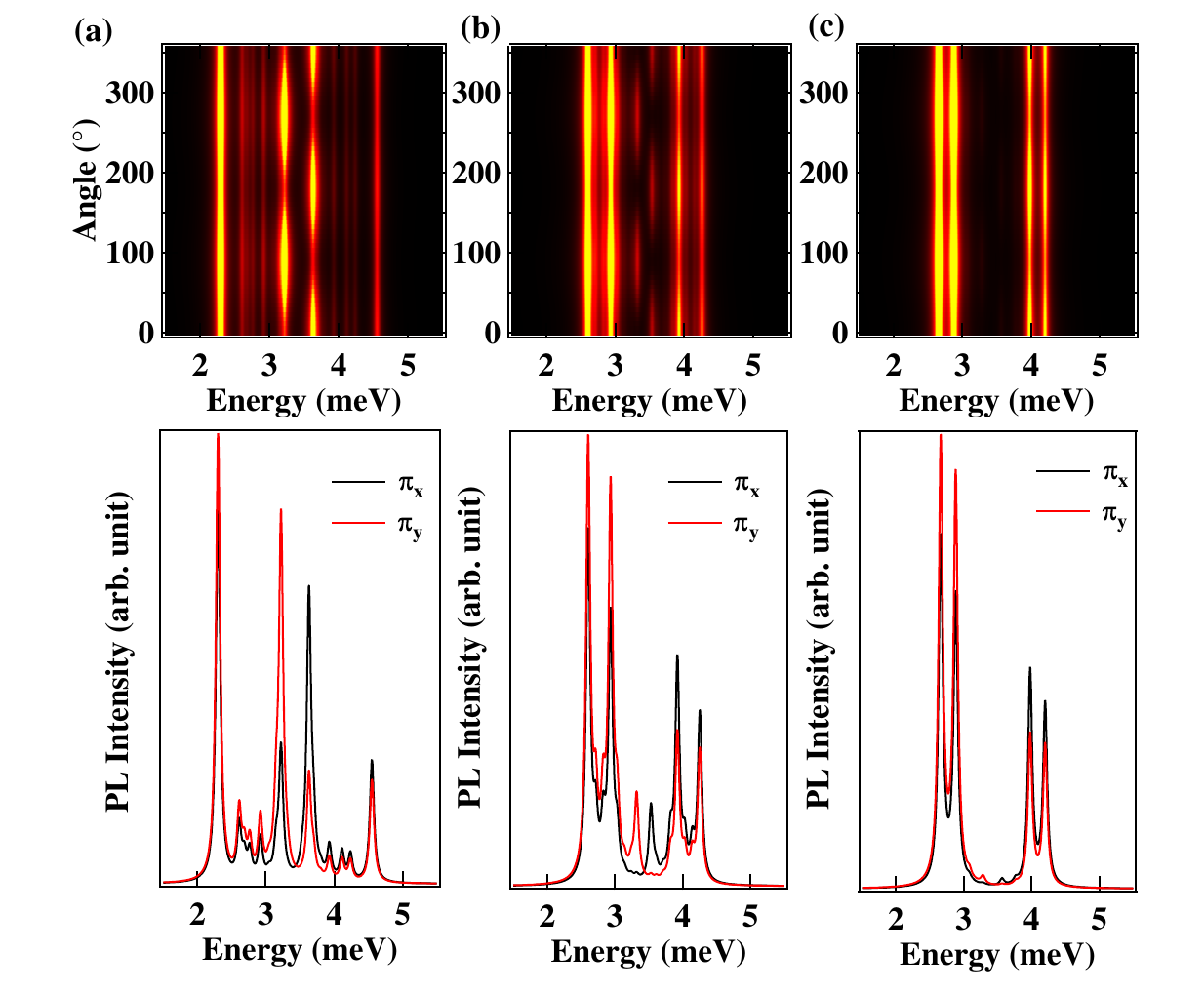} 
    \caption{Calculated zero magnetic field linearly polarized PL spectra of a QD containing 2 Co atoms with (a) the parameters listed in table \ref{TablePar}, (b) parameters of table \ref{TablePar} and I$_{hCo(2)}$=75$\mu eV$ and (c) parameters of (b) with D$_{0(1)}$=D$_{0(2)}$=-3.5 meV. For all spectra T$_{eff}$=20 K. A line broadening with a Lorentzian of full width at half maximum of 75 $\mu$eV is included in the spectra. In the PL intensity maps (top panels) the angle is measured from the [100] axis.} 
    \label{fig:model0B} 
\end{figure}

The band Hamiltonian, $\mathcal{H}_{bd}=E_g+\mathcal{H}_{vb}$, stands for the energy of the electron, $E_g$, and the heavy-holes and light-holes energies $\mathcal{H}_{vb}$. We consider here the four lowest energy holes states $\vert J_h,J_{hz}\rangle$ with J$_h$=3/2. A general form of Hamiltonian describing the influence of shape or strain anisotropy on the valence band structure can be written in the basis ($|\frac{3}{2},+\frac{3}{2}\rangle,|\frac{3}{2},+\frac{1}{2}\rangle,|\frac{3}{2},-\frac{1}{2}\rangle,|\frac{3}{2},-\frac{3}{2}\rangle$) as:

\begin{equation}\label{Hvbm}
\mathcal{H}_{vb} = \left(
\begin{array}{cccc}
0                                &-Q                                           &R                                   &0\\
-Q^*                             &\Delta_{lh}                                  &0                                   &R\\
R^*                              &0                                            &\Delta_{lh}                         &Q\\
0                                &R^*                                          &Q^*                                 &0\\
\end{array}\right)
\end{equation}

\noindent with $R=\delta_{xx,yy}-i\delta_{xy}$ and $Q=\delta_{xz}-i\delta_{yz}$. Here, $\Delta_{lh}$ is the splitting between heavy-hole (hh) and light hole (lh) which is controlled in QDs both by the in-plane biaxial strain and the confinement. R describes the heavy-hole/light-hole mixing induced by an anisotropy in the QD plane $xy$ and Q takes into account an asymmetry in the plane containing the QD growth axis $z$. The reduction of symmetry can come from the shape of the QD (Luttinger Hamiltonian) or the strain distribution (Bir and Pikus Hamiltonian). R is usually written in the form $R=\rho_s/\Delta_{lh}e^{-2i\theta_s}$ where $\rho_s/\Delta_{lh}$ describes the amplitude of the band mixing and $\theta_s$ is the direction of the main anisotropy responsible for the band mixing measured from the [100] axis.
 
Considering only an in-plane anisotropy (Q=0), it follows from (\ref{Hvbm}) that the band mixing couples the heavy-holes $J_z=\pm3/2$ and the light-holes $J_z=\mp1/2$ respectively. For such mixing, the short range exchange interaction, which can be written in the form $2/3\delta_0(\overrightarrow{\sigma}.\overrightarrow{J})$, couples the two bright excitons. A deformation in a vertical plane (Q term) can couple the heavy-holes $J_z=\pm3/2$ and the light-holes $J_z=\pm1/2$ respectively. In this case, the short range electron-hole exchange interaction couples $|+1\rangle$ and $|+2\rangle$ exciton on one side and $|-1\rangle$ and $|-2\rangle$ exciton on the other side.

Using the Hamiltonian of the excited state $\mathcal{H}_{X,2Co}$ and the Hamiltonian of the ground  state 

\begin{eqnarray}
\mathcal{H}_{2Co}=\sum_i(\mathcal{H}_{s,Co(i)}+g_{Co(i)}\mu_B\vec{B}.\vec{S_i})
\end{eqnarray}

\noindent we can compute the PL spectrum of QDs containing two Co atoms. The occupation probability of the exciton-2Co levels is described by an effective spin temperature T$_{eff}$ and the optical transition probabilities are obtained calculating the matrix elements $\vert\langle S_{z,i}\vert X,S_{z,i}\rangle\vert^2$.

\begin{figure}[hbt]
    \centering
    \includegraphics[width=1.0\linewidth]{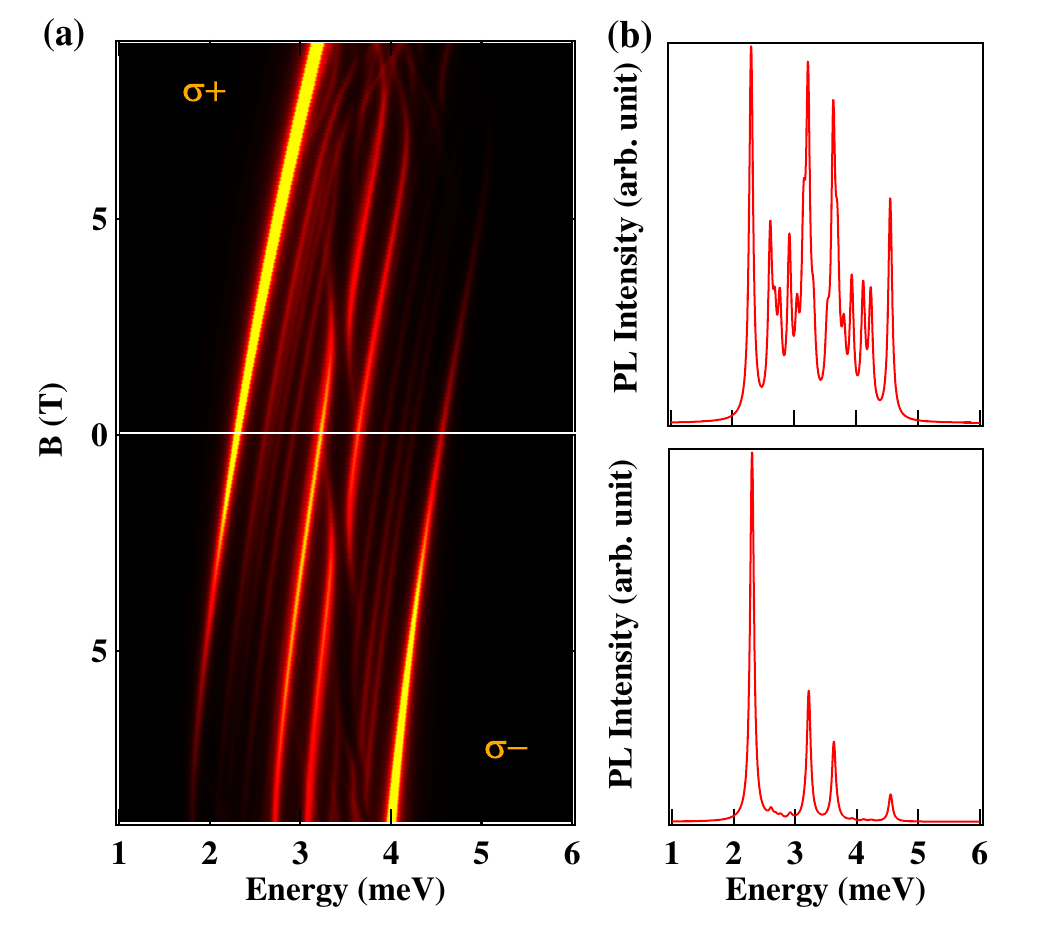} 
    \caption{(a) Calculated magnetic field dependence of the PL of a QD containing two Co atoms with the parameters of Table \ref{TablePar} and T$_{eff}$=20K. (b) Calculated zero field spectra for two effective spin temperatures, T$_{eff}$=40K (top) and T$_{eff}$=10K (bottom). A line broadening with a Lorentzian of full width at half maximum of 75 $\mu$eV is included in the spectra.} 
    \label{fig:modelB} 
\end{figure}

Calculated linearly polarized spectra at zero magnetic field are presented in Fig.~\ref{fig:model0B}. Unless specified, the parameters are those listed in table \ref{TablePar}. The change from dot to dot of the relative position of the inner and outer lines, the appearance of low intensity lines and the appearance of linear polarization in the center of the PL spectrum are well explained by the model.  

A calculated longitudinal magnetic field dependence is also presented in Fig.\ref{fig:modelB}. The exchange interaction terms (see table \ref{TablePar}) are chosen to reproduce the splitting of the four main lines of QD3 and the other parameters are typical for single Co \cite{Kobak2018} and individual CdTe/ZnTe QDs \cite{Besombes2023}. Even if some details are not described, this simplified spin-effective model reproduces accurately the main experimental observations under magnetic field (see Fig.~\ref{fig:MagnetoPL1}(a) for QD3). In particular the overall  evolution under magnetic field with its characteristic intensity distribution is well described. The behavior of the weaker intensity lines and the increase of their contribution with the increase of an effective spin temperature T$_{eff}$ are also well reproduced (Fig.~\ref{fig:MagnetoPL1}(b)). Additional anticrossings or weak intensity line can appear in the spectra of some of the dots (see QD2) but the overall behavior remains the same. Such deviation may result from a further reduction in the symmetry due to the position of the magnetic atoms or disorientation of the axis of local strain \cite{Bhattacharjee2007} but these parameters cannot be accurately and independently adjusted for the two atoms.

\subsection{Carriers-induced spin-spin correlation}

A remarkable feature of 2Co-doped QDs is that it provides a carrier mediated coupling between both distant Co spins. At zero magnetic field, the exchange interaction with the exciton, dominated by the coupling with the hole, results in two ground states $\vert\Downarrow_h,\uparrow_e\rangle\vert\frac{3}{2}\rangle_{(1)}\vert\frac{3}{2}\rangle_{(2)}$ or $\vert\Uparrow_h,\downarrow_e\rangle\vert-\frac{3}{2}\rangle_{(1)}\vert-\frac{3}{2}\rangle_{(2)}$  with a ferromagnetic alignment of the two Co spins.

\begin{figure}[hbt]	
    \centering
    \includegraphics[width=1.0\linewidth]{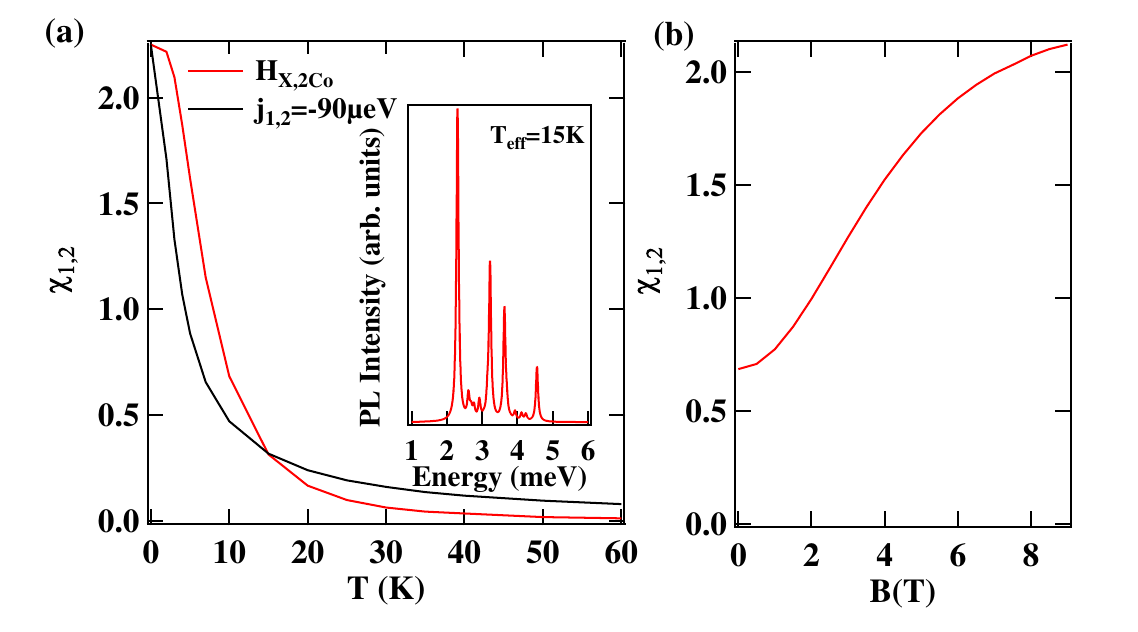} 
    \caption{Correlation function $\chi_{1,2}$ of the two Co spins in a QD. (a) Temperature dependence for the model presented in the text $\mathcal{H}_{X,2Co}$ calculated with the parameter of Table \ref{TablePar} and compared with the spin-spin correlation function for a Heisenberg model for two spins S = 3/2 with j$_{1,2}$=-90$\mu$eV. The inset show a calculated PL spectra with the T$_{eff}$=15K corresponding to $\chi_{1,2}$=0.032 which can be reproduced with a Heisenberg coupling j$_{1,2}$=-90. (b) Longitudinal magnetic field dependence of the spin-spin correlation function at T$_{eff}$=10K calculated with $\mathcal{H}_{X,2Co}$ and parameters of Table \ref{TablePar}.}
    \label{fig:Cor} 
\end{figure}

Even in absence of spin polarized carriers, the thermalization of the Co spins during the lifetime of the exciton induces a positive spin-spin correlation as would be produced by a ferromagnetic exchange between the Co spins. This correlation is observed at zero field in the intensity ratio of the high and low energy lines which reflects the increase of the probability to have parallel Co spins:  $\vert-\frac{3}{2}\rangle_{(1)}\vert-\frac{3}{2}\rangle_{(2)}$  for a $\sigma+$ exciton and $\vert+\frac{3}{2}\rangle_{(1)}\vert+\frac{3}{2}\rangle_{(2)}$ for a $\sigma-$ exciton. The spin correlation function of the two Co atoms is given by:

\begin{eqnarray}
\chi_{1,2}(k_BT)=\langle \vec{S_1}.\vec{S_2} \rangle= \sum_X P_X(k_BT)\langle\psi_X\vert \vec{S_1}.\vec{S_2} \vert \psi_X\rangle
\end{eqnarray}

\noindent where P$_X=e^{-E_X/K_BT}/\Sigma_ne^{-E_n/K_BT}$ is the Boltzmann occupation function of the emitting states. Perfectly aligned ferromagnetically coupled spins will give $\chi_{1,2}$=9/4.

We can calculate $\chi_{1,2}$ as a function of temperature for the exciton mediated coupling, controlled by $\mathcal{H}_{X,2Co}$, for a given set of parameters and for a direct exchange j$_{1,2}\vec{S_1}.\vec{S_2} $ between Co spins. These curves are presented in Fig.~\ref{fig:Cor}(a). The parameters for the exciton model are those of Table~\ref{TablePar} and for the Heisenberg coupling $j_{1,2}$=-90$\mu eV$. The two correlations curves present a similar temperature behavior thought $j_{1,2}$ cannot be adjusted to perfectly match the exciton induced correlation both at low and high temperature.

To reproduce the intensity ratio which is for instance observed for QD3 in Fig.\ref{fig:PLPi}, an effective spin temperature T$_{eff}$ $\approx$ 25 K is used. This gives a correlation function $\chi_{1,2}$=0.1. This measured spin correlation can be interpreted as a Zener type ferromagnetism controlled at the nano-scale by individual carriers \citep{Dietl2000}.

Excitation conditions which minimizes the heating by high energy carriers can enhance the correlation between the two localized spins. This correlation can be further improved by the application of a longitudinal magnetic field that induces a Zeeman splitting of each atom and enhance their alignment (see Fig.~\ref{fig:Cor}(b)). The injection of spin polarized carriers that can produce a progressive orientation of the two Co spins in the exchange field of the exciton can also significantly enhance their correlation.

\section{Optical orientation and dynamics of the spin of two Co atoms.}

\subsection{Optical orientation}

\begin{figure}[hbt]
    \centering
    \includegraphics[width=1.0\linewidth]{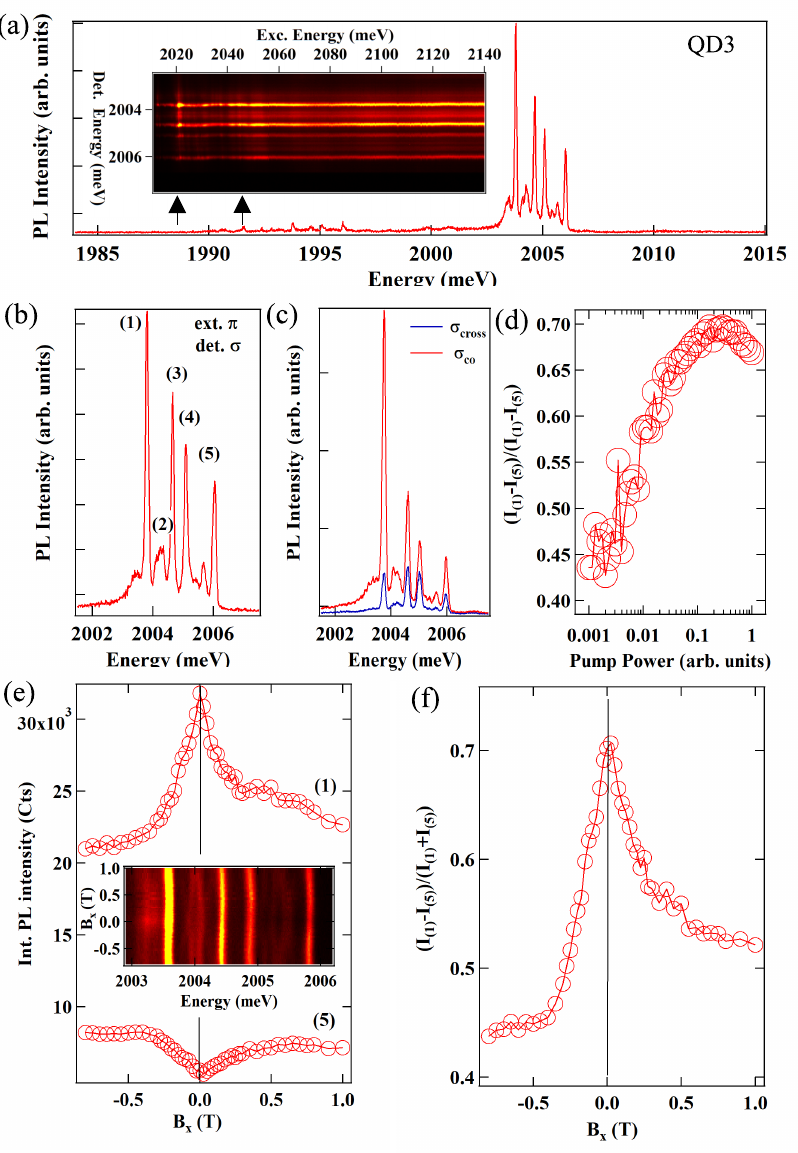} 
    \caption{(a) Large band PL spectra of QD3. The inset shows an intensity map of the PLE of QD3. Arrows point the two excitation energies used in the experiments discussed in the text. (b) Circularly polarized PL spectra of QD3 for a linearly polarized excitation at 2050 meV. (c) Co and Cross circularly polarized PL spectra of QD3 for a circularly polarized excitation at 2050 meV. (d) Excitation power dependence of the PL intensity ratio of the high energy and low energy lines of QD3 (I$_{(1)}$-I$_{(5)}$)/(I$_{(1)}$+I$_{(5)}$). (e) Transverse magnetic field dependence of the normalize PL intensity of the high and the low energy lines of QD3 for a $\sigma_{Co}$ excitation/detection. The inset shows the transverse magnetic field dependence of the PL intensity map. (f) Magnetic field dependence of the PL intensity ratio of the high energy and low energy lines of QD3 (I$_{(1)}$-I$_{(5)}$)/(I$_{(1)}$+I$_{(5)}$).} 
    \label{fig:refQD} 
\end{figure}

Spin polarized excitons can be optically injected in the dots through a circularly polarized excitation on excited states (see the PLE spectrum of QD3 in the inset of Fig.~\ref{fig:refQD}(a)). We can first notice that under these excitation conditions, the PL is strongly co-circularly polarized with the excitation (Fig.~\ref{fig:refQD}(c)). This shows that the spin polarization of the exciton is well conserved during its relaxation and during its lifetime in the QD ground state.  

More interestingly, the relative intensities of the PL lines of the exciton-2Co complex strongly changes when the polarization of the excitation is changed from linear to circular. This is presented in Fig.~\ref{fig:refQD}(b) and (c) for an excitation at 2050 meV. For a linearly polarized excitation and a circular detection the intensity ratio of the high and low energy lines reflects a thermalisation of the two Co spins in the exchange field of the exciton. 

For a $\sigma_{Co}$ excitation/detection, this intensity ratio is significantly modified and most of the PL comes from the low energy line. As each emission line corresponds to a given spin state of the 2 Co, M$_z$, this shows that the injection of spin polarized excitons creates a non-equilibrium distribution of the Co spin states which can be described by a lower effective spin temperature. In the excitation configuration presented in Fig.~\ref{fig:refQD}(c), the intensity ratio of the high an low energy lines (I$_{(1)}$-I$_{(5)}$)/(I$_{(1)}$+I$_{(5)}$)=0.7 can be reproduced with T$_{eff}$=15K (see inset of Fig.~\ref{fig:Cor}(a)) giving an enhanced correlation function $\chi_{1,2}\approx$ 0.32.

The change in the intensity distribution between the $\pi$ and the $\sigma$ polarization configurations reflects an optical pumping of the two Co spins for a $\sigma$ excitation {\it i.e.} a progressive orientation of the two Co spins in the exchange field of successively injected spin polarized excitons. This cumulative effect is confirmed by the excitation power dependence of the high energy/low energy lines intensity ratio (I$_{(1)}$-I$_{(5)}$)/(I$_{(1)}$+I$_{(5)}$) presented in Fig.~\ref{fig:refQD}(d). The increase of the intensity ratio with the increase of the excitation power results from an enhancement of the efficiency of pumping at high power. A saturation and a slight decrease of this intensity ratio is however observed at high excitation power. It can result from a heating of the spins by (i) the non-resonantly carriers created in the wetting layer and/or (ii) the interaction with non-equilibrium phonons generated during the relaxation of carriers \cite{Besombes2008,Tiwari2020}. 

The high energy/low energy lines intensity ratio in $\sigma_{Co}$ configuration also strongly depends on a transverse magnetic field. This is presented in Fig.~\ref{fig:refQD}(e) for a circularly polarized excitation of QD3 at 2050 meV (see Fig.~\ref{fig:refQD}(a)). As the transverse field is applied, a decrease of the intensity of the low energy line and an increase of the intensity of the high energy line are simultaneously observed. The corresponding normalized intensity ratio is presented in Fig.~\ref{fig:refQD}(f). The reduction of this intensity ratio corresponds to a decrease of the spin polarization of the two atoms. The depolarization curve in transverse magnetic field presents a slight asymmetry.

The opposite evolution of the intensity of the low- and high-energy lines shows that the transverse magnetic field dependence is not simply the result of a reduction in exciton spin polarization. Such a reduction would lead to a change in the polarization rate of all the exciton-2Co lines, but not to a change in the intensity distribution between the lines. The transverse magnetic field dependence results indeed from a reduction in the optical pumping efficiency. As observed in Fig.~\ref{fig:refQD}(f), a transverse field of about 0.4 T is required to significantly erase the optical pumping.

The effect of the weak magnetic field can be neglected in the excited state where the Zeeman energy of the atoms remains much weaker than the exciton exchange field oriented along the growth axis of the QD. Depolarization should then be controlled by the precession of the two Co spins in the transverse magnetic field. In the QD ground state, this precession is limited by the fine structure of Co atoms induced by bi-axial strain. However, as all the 2Co spins ground states with M$_z$=$\pm$3 and M$_z$=0 are degenerate, a weak mixing term can induce a coherent coupling between them. 

The asymmetry in the depolarization can come from the presence of in-plane anisotropy in the magnetic atoms fine-structure (E term). Depending on the in-plane orientation of this anisotropy, the response of the energy levels of the atoms to the transverse field can depend on its orientation. This has been observed in the time domain in the case of individual Mn atoms \cite{Lafuente2015}. Probing in detail this anisotropy would require further magneto-optic measurements with a change of the direction of the transverse magnetic field in the plane of the QD which could not be performed in the available set-up. The influence of the E term on 2Co dots will be discussed more in details in Sec. V.

The observation of an optical orientation shows that, even in the presence of the strain induced magnetic anisotropy D$_0$, some spin-flip of the magnetic atoms can occur during the lifetime of an exciton. However this experiment does not permit to identify the main channel of spin-flip responsible of the pumping and their time-scale.

\subsection{Spin dynamics revealed in auto-correlation measurements}

Observing the time evolution of the PL intensity of a magnetic dot can be used to probe the spin fluctuations of the Co atoms in the presence of the exciton. This spin dynamics is responsible for the observed optical orientation. PL intensity fluctuations and their characteristic time can be revealed by the auto-correlation signal of the emission of a given line of the dot.

\begin{figure}[hbt]
    \centering
    \includegraphics[width=1.0\linewidth]{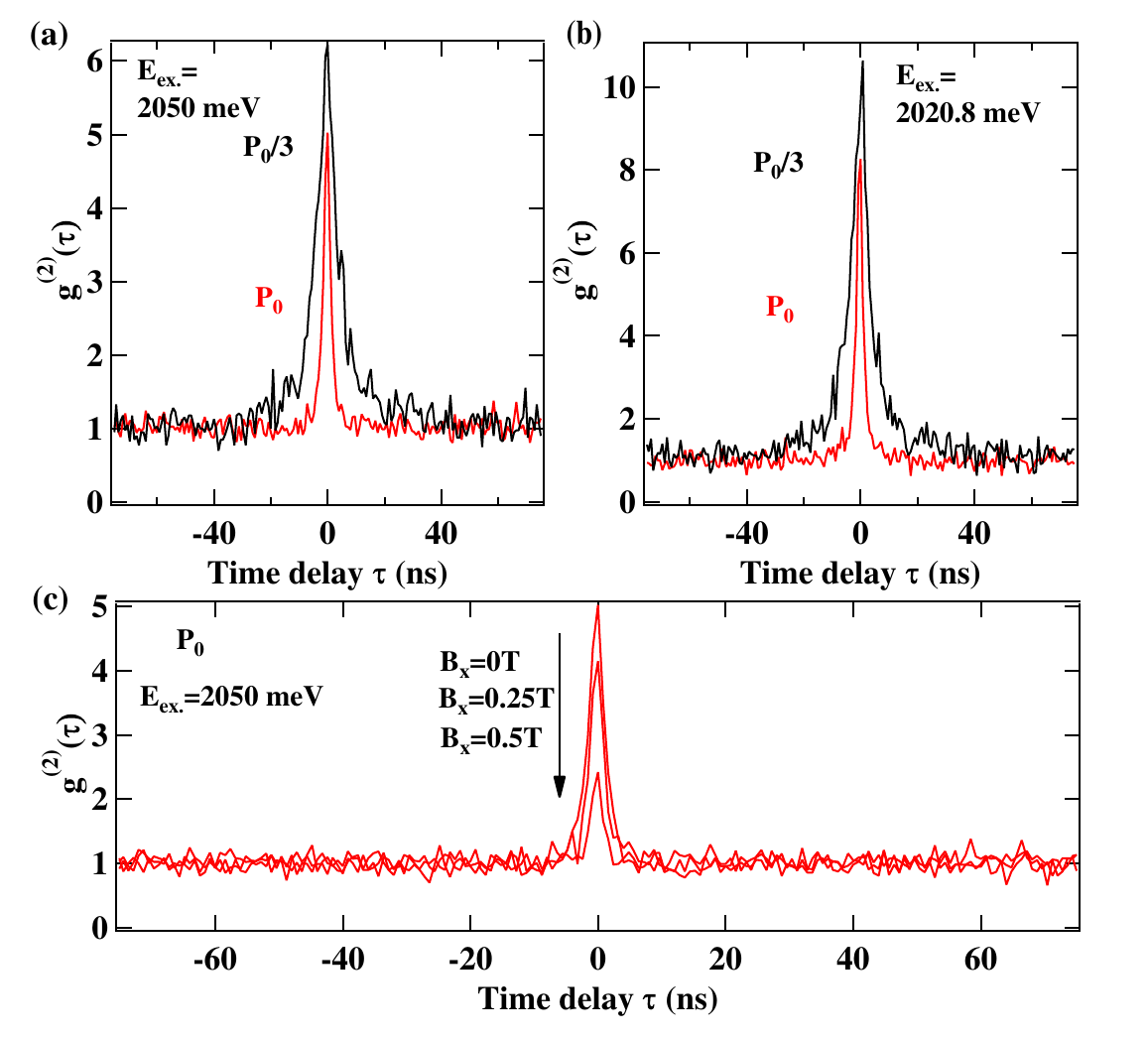} 
    \caption{Auto-correlation of the $\sigma_{Co}$ PL intensity of the low energy line (5) of QD3 for two different energies of excitation, (a) 2050 meV and (b) 2020.8 meV and in each case for two excitation powers, P$_0$ and 3P$_0$. (c) Transverse magnetic field dependence of the auto-correlation signal for an excitation at 2050 meV and an excitation power P$_0$.} 
    \label{fig:autocor} 
\end{figure}

To estimate the auto-correlation signal we used a HBT setup with a time resolution of about 700 ns \cite{Lafuente2016}. In this experiment, the detection of a first circularly polarized photon from a given line indicates that the system is in a particular spin state of the two atoms, M$_z$. The probability of delayed detection of a second photon with the same energy and polarization depends on the probability of conservation of the spin state and on the excitation rate. This photons’ coincidence measurement is a good approximation of the auto-correlation of the PL intensity as far as the considered time delay remains smaller than the inverse of the photon collection rate limited in these experiments to a few kHz by the collection efficiency of a single QD emission.

An auto-correlation signal is presented in Fig.~\ref{fig:autocor} for a detection of the intensity of the low energy line of QD3, under excitation at 2050 meV and at zero magnetic field. Because of the limited time resolution of the experimental setup and the short lifetime of the exciton (around 250 ps in CdTe/ZnTe QDs), the antibunching characteristic of a single-photon emitter is not observed. However, a photon bunching with an amplitude of about 6 and a characteristic width in the tens of nanoseconds range is obtained. This clearly shows intermittent QD emission with the amplitude of the bunching proportional to the OFF/ON ratio of the selected PL line. 

As presented in Fig.~\ref{fig:autocor}, the amplitude of the bunching depends on the excitation energy and slightly increases for a resonant laser on a lower energy and sharper excited state (Fig.~\ref{fig:autocor}(b)). The width of the bunching reveals the timescale of the intensity fluctuations. For both excitation energies, the speed of fluctuations increases with the increase of the excitation intensity. 

Charge fluctuations are unlikely in the studied magnetic QDs as only the neutral exciton is observed in the PL spectra (see Fig.~\ref{fig:refQD}(a)). The measured fluctuations timescale is controlled both by the fluctuations in the excited state, when the QD is occupied by an exciton and by the spin dynamics in the ground state when the QD is empty. The reduction of the linewidth of the bunching with the increase of the excitation intensity could traduce a faster spin dynamics in the excited state when an exciton is present in the dot than in the empty dot. It is also possible that the high-energy and high-intensity optical excitation generates non-equilibrium acoustic phonons that can induce spin flips of the Co spins \cite{Tiwari2020}. The two processes probably contribute to the acceleration of the spin dynamics and this experiment does not permit to discriminate the main channels of spin relaxation. 

As it was observed in the optical orientation experiment, a weak transverse magnetic field also significantly affects the auto-correlation signal (Fig.~\ref{fig:autocor}(c)) with both a decrease of the amplitude and of the width of the photon bunching peak. As the exciton-2Co exchange interaction is much larger than the Zeeman energy, the transverse field dependence of the measured dynamics is a consequence of the field induced mixing of the Co spins states in the empty dot.

\section{Resonant photo-luminescence of QDs doped with two Co atoms}

To avoid heating of spins by the injection of high-energy carriers or carrier induced spectral diffusion, resonant optical excitation is generally preferred for optical control of individual spins or generation of high quality individual photons with QDs \cite{DeSantis2017}. Resonant excitation combined with the detection of quasi-resonant PL has been successfully used for the initialization of the spin of individual Mn$^{2+}$, Cr$^{2+}$ atoms and hole-Cr$^+$ complex \citep{Lafuente2017,Tiwari2020,Tiwari2022}.

We use here resonant optical excitation of the neutral exciton lines to analyze the spin transfer mechanism among the exciton-2Co complex. This technique also enables to study the extent to which the two Co spins can be optically initialized by resonant optical pumping. 

For this experiment we select a dot, QD3, that is mainly neutral ({\it i.e.} no significant contribution of the charged excitons is observed in the PL, see Fig.~\ref{fig:refQD}). The resonant PL and the spin dynamics in QD2 are unlikely to be affected by any charge fluctuations in the dot. This dot also presents a large splitting which permits to independently address each emission line with a resonant optical excitation. Quasi-resonant PL is studied in QD3 by scanning a tunable laser on the high energy side of the exciton-2Co PL spectra while detecting the emission of the low energy line (1) or the central line (3) (see Fig.~\ref{fig:refQD} for the labeling of the lines).

\subsection{Absorption measured in the acoustic-phonon side-band}

\begin{figure}[hbt]
    \centering
    \includegraphics[width=1.0\linewidth]{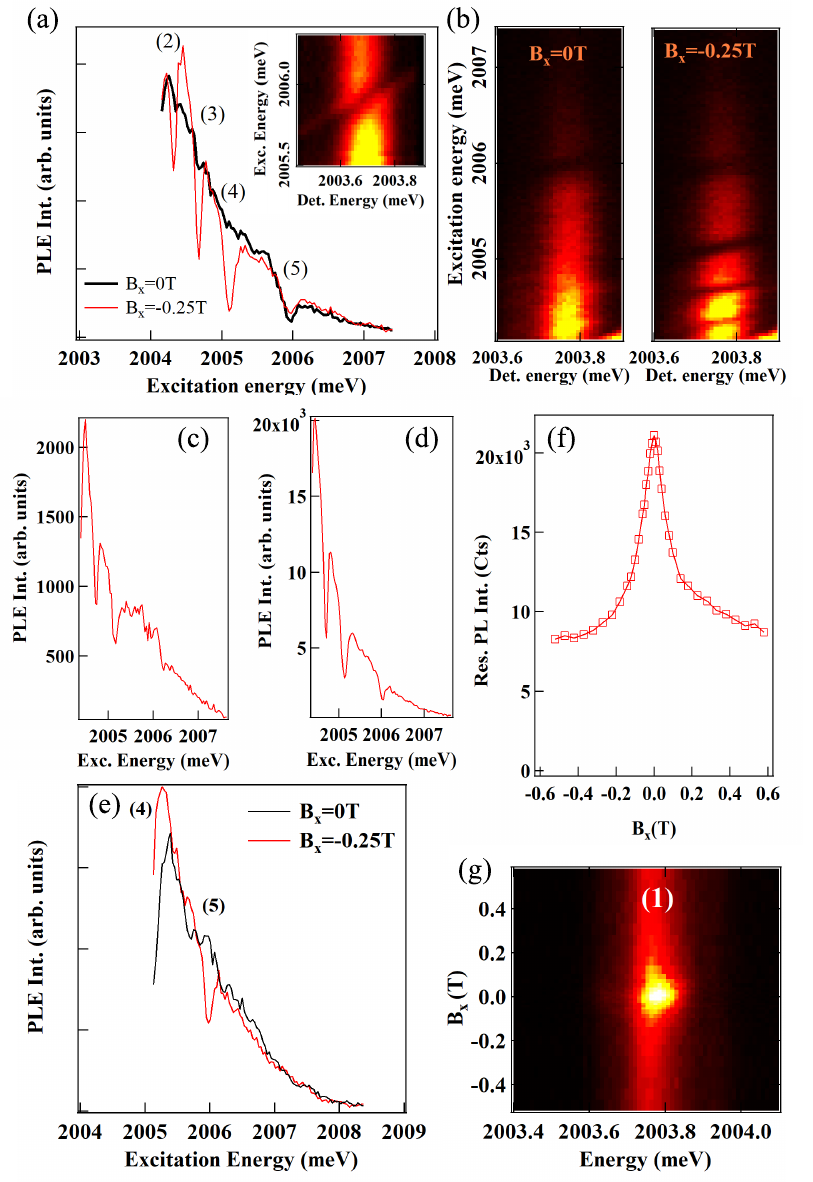} 
    \caption{(a) PLE intensities detected on line (1) of QD3 as a function excitation energy at zero magnetic field and under a transverse field B$_x$=-0.25T. The inset shows a detail of the PL intensity map of line (1) at B$_x$=0T for an excitation around (5). (b) PL intensity map of the low energy line (1) of QD3 for a tunable quasi-resonant excitation at zero magnetic field (left) and under a transverse field B$_x$=-0.25T (right) for a cross-linear excitation/detection configuration. (c) and (d) PLE intensity detected on line (1) of QD3 under B$_x$=-0.25T for a cross-circular excitation/detection and for a co-circular excitation/detection configuration respectively. (e) Integrated PL intensity of line (3) of QD3 as a function excitation energy at zero magnetic field and under a transverse magnetic field B$_x$=-0.25T. (f) Transverse magnetic field dependence of the quasi-resonant $\pi_{cross}$ PL intensity for a detection on line (1) and an excitation on line (4). (g) Corresponding PL intensity map of line (1).} 
    \label{fig:PLE}  
\end{figure}

A PLE spectra obtained in $\pi_{cross}$ excitation/detection configuration for a detection on the low energy line (1) of QD3 is presented in Fig.~\ref{fig:PLE} (a) and the corresponding quasi-resonant PL intensity map in Fig.~\ref{fig:PLE} (b). A quasi-resonant PL is observed for excitation very close to the emission line and up to a few milli-electronvolts above it. A progressive decrease in PL intensity occurs with increasing the excitation energy. This side-band is characteristic of the absorption in the acoustic phonon band of the observed emission line \cite{Besombes2001,Favero2003}. 

Surprisingly, at zero magnetic field, the quasi-resonant PL spectra does not reveal any excitation transfer. Excitation transfer usually appears as an increase of the PL which becomes larger than the acoustic phonon background for resonant excitation on some specific excitonic line. This was for instance observed in the case of Mn$^{2+}$, Cr$^{2+}$ or hole-Cr$^{+}$ in similar QDs and permit to identify the dominant spin relaxation channels of the exciton / magnetic atoms complex \citep{Lafuente2017,Tiwari2020,Tiwari2022}. 

The absence of any additional PL signal superimposed on the acoustic phonon background shows that in the case of the 2Co system there is no efficient excitation transfer during the exciton lifetime. This can either come from (i) an efficient resonant optical pumping which prevents the resonant absorption or (ii) from an absence of significant spin relaxation for a resonantly created exciton.

At zero magnetic field, the only feature observed in the PLE is a dip that appears in the absorption background when the excitation laser is scanned around the high-energy line (5). The exact energy position of the dip follow the excitation laser as it crosses the exciton line broadened by spectral diffusion (inset of Fig.~\ref{fig:PLE}(a)). Dips are not observed for an excitation on the other lines. This shows that this reduction in PL intensity for an excitation on (5) does not result from saturation of absorption by a direct resonant excitation of a transition. Rather, such saturation would produce a dip for excitation on each of the excitonic lines.
 
As the low- and the high-energy lines correspond to the same spin states of the two Co atoms, {\it i.e.} $\vert\frac{3}{2}\rangle_{(1)}\vert\frac{3}{2}\rangle_{(2)}$ or $\vert-\frac{3}{2}\rangle_{(1)}\vert-\frac{3}{2}\rangle_{(2)}$ for a $\pi_{cross}$ excitation/detection, the reduction in PL intensity reflects a decrease of occupancy probability of these two spin states. This reduction is induced by a resonant optical pumping: resonant absorption followed by a spin-flip in the exciton-2Co system empties the states that are resonantly excited.

\subsection{Transverse magnetic field dependence of the absorption of the acoustic phonon side-band}

The behavior of the absorption in the acoustic phonon band significantly changes when a weak transverse magnetic field is applied. This is illustrated in Fig.~\ref{fig:PLE}(a) and (b) for a detection on the low energy line (1) of QD3: under a transverse field, additional dips appear in the side-band for resonant excitation of some of the higher energy excitonic lines. For a transverse field B$_x$=-0.25T, dips are clearly observed for an excitation on the higher energy lines (3) and (4) and also for an excitation of the weaker intensity PL line (2).

Quasi-resonant PL with a similar behavior in transverse magnetic field is also observed for a circularly polarized excitation in the acoustic phonon band. In this case, the PL is mainly co-circularly polarized with the excitation (Fig.~\ref{fig:PLE}(c) and (d)). This confirms the good stability of the spin of the carriers coupled with the spins of the two magnetic atoms.  

For a detection on the central line (3) (see Fig.~\ref{fig:PLE}(e)), at zero magnetic field, an unperturbed acoustic phonon absorption background is observed except for an excitation on (4). When a weak transverse magnetic field is applied, a dip appears in the phonon side-band for excitation around the high energy line (5) showing that a transverse field creates a link between the outer and the inner lines.

The amplitude of the dips in the acoustic phonon bands are controlled by the transverse magnetic field. This is illustrated in Fig.~\ref{fig:PLE}(f) and (g) for a detection on (1) and a resonant excitation on (4). When the transverse magnetic field is increased, a progressive decrease of the resonant PL is observed. A minimum of resonant PL intensity is reached around B$_x$=0.5 T. As it was observed for the optical orientation under injection of spin polarized carriers, a slight asymmetry in the transverse magnetic field dependence is observed. It likely arises from the orientation of the in-plane strain anisotropy at the origin of the Co atoms fine structure. A full width at half maximum on the dependence of the transverse field can nevertheless be defined with a value around 0.16 T.

This transverse magnetic field dependence of the dips confirm that they do not arise from a saturation of the absorption of another line of the dot as the oscillator strength of a transition is not expected to be significantly affected by such weak magnetic field. However, these experiments demonstrate that a transverse magnetic field creates a link between the optical transitions that do not share the same 2Co spins ground states. For instance, transition (1) is associated with $\vert+\frac{3}{2}\rangle_{(1)}\vert+\frac{3}{2}\rangle_{(2)}$ or $\vert-\frac{3}{2}\rangle_{(1)}\vert-\frac{3}{2}\rangle_{(2)}$ and transitions (3) and (4) with $\vert-\frac{3}{2}\rangle_{(1)}\vert+\frac{3}{2}\rangle_{(2)}$ or $\vert+\frac{3}{2}\rangle_{(1)}\vert-\frac{3}{2}\rangle_{(2)}$. The corresponding optical transitions are independent at zero field.

Whereas a large effect of the transverse field is observed for a resonant excitation on the intermediate lines (2), (3), (4), the dip on the high energy line is not significantly affected in the investigated field range. This can come from the fact that to observe a signal for an excitation in the acoustic phonon band at the energy of the high energy line, the excitation intensity is large. It likely saturates the absorption of the high energy line when on resonance and prevents the effect of the weak transverse field which only act on the empty dot.

\subsection{Discussion on the transverse magnetic field dependence of the resonant PL}

For the quasi-resonant PL experiments discussed here, the low-energy line is continuously excited by absorption in the acoustic phonon band. This absorption is not spin-selective and all the optical transitions of the dots are simultaneously excited, though with slightly different probabilities, depending on their distance to the resonant laser. The excitation in the acoustic-phonon band does not contribute to an optical pumping of the two Co spins. In the case of a linearly polarized excitation, the excitons are not spin-polarized, nor can they induce optical Co spins orientation. The weak transverse field used in this experiment does not significantly affect the energy levels of the exciton-2Co complex. 

In the ground state the four spin states with the largest occupation probability $\vert\pm3/2\rangle_{(i)}$ are initially degenerate at zero magnetic field and they can be coherently coupled by any weak mixing terms. The spin states $\vert\pm1/2\rangle_{(i)}$ are pushed to high energy by the bi-axial strain term D$_0$. The presence of in-plane strain anisotropy induces a fine structure term $E$ which couples spin states $\vert+3/2\rangle_{(i)}$ and $\vert-1/2\rangle_{(i)}$ on one side and $\vert-3/2\rangle_{(i)}$ and $\vert+1/2\rangle_{(i)}$ on the other side. The coupling is controlled by the ratio $E/D_0$. A magnetic field applied in the plane of the QD induces a coupling of the spin states $\vert\pm3/2\rangle_{(i)}$ and $\vert\pm1/2\rangle_{(i)}$ respectively. Combined with the presence of a $E$ term, these two mixing induces a coupling between the $\vert\pm3/2\rangle_{(i)}$ spin states of each atom which are initially degenerate. The resulting mixing of the four ground states depends on the amplitude of the applied transverse magnetic field. 

A dip in the acoustic phonon band of the low energy line (1) for a resonant excitation of a higher energy line in $\pi_{cross}$ excitation-detection configuration means a decrease of the population of the spin state corresponding to the low energy line, $\vert\pm3/2\rangle_{(1)}\vert\pm3/2\rangle_{(2)}$. This shows that under a transverse magnetic field, a resonant excitation of the lines associated with $\vert\pm3/2\rangle_{(1)}\vert\mp3/2\rangle_{(2)}$ (lines (3) or (4)) introduces a channel of pumping for the spin states $\vert\pm3/2\rangle_{(1)}\vert\pm3/2\rangle_{(2)}$.

This channel of pumping is induced by the combined action of the E term and the transverse magnetic field. For instance in the presence of a transverse field, the ground state $\vert+3/2\rangle_{(1)}$+$\vert-3/2\rangle_{(2)}$ corresponding to line (3) of QD3, becomes ($\vert+3/2\rangle_{(1)}$+$\epsilon_{(1)}$ $\vert-3/2\rangle_{(1)}$)+($\vert-3/2\rangle_{(2)}$+$\epsilon_{(2)}$ $\vert+3/2\rangle_{(2)}$). With this mixing, a resonant optical excitation on line (3) can also excite the dot if the spins are initially in the state  $\vert+3/2\rangle_{(1)}$+$\vert+3/2\rangle_{(2)}$. This introduces an efficient channel of optical pumping as an absorption on the central line and a recombination projects the two Co spins on the states $\vert\pm3/2_{(1)};\mp3/2_{(2)}\rangle$ different from the initial 2Co spins ground state. 

A similar scenario explains that the resonant excitation of some transitions with weak PL signal efficiently decrease the intensity of the low energy line. Let's consider for instance the optical transition associated with the 2Co spins state $\vert-3/2\rangle_{(1)}$)+$\vert-1/2\rangle_{(2)}$. It corresponds to the lowest energy level containing a $S_z=\pm1/2$ component (see Fig.~\ref{fig:scheme}). If we take into account the mixing induced by the fine structure term $E$ it becomes ($\vert-3/2\rangle_{(1)}$)+($\vert-1/2\rangle_{(2)}$+$\xi_{(2)}$ $\vert+3/2\rangle_{(2)}$). Under a transverse field this state becomes ($\vert-3/2\rangle_{(1)}$+$\epsilon_{(1)}$ $\vert+3/2\rangle_{(1)}$)+($\vert-1/2\rangle_{(2)}$+$\xi_{(2)}$ $\vert+3/2\rangle_{(2)}$). A resonant optical excitation of the corresponding line can also excite the dot if the spins are initially in the state $\vert+3/2\rangle_{(1)}$+$\vert+3/2\rangle_{(2)}$. This introduces a channel of pumping controlled by the transverse field and the initial strain induced coupling.

\section{Conclusion}

To conclude, we demonstrated the possibility to optically access the spin states of two Co atoms in a semiconductor host. Magneto-optic measurements and modeling of QDs containing two Co atoms show that a large diversity of spectra can be obtained. PL spectra depend on the relative position of the two atoms inside the dot and on the strain state at each magnetic atom location. In most of the dots, the emission is dominated by four main lines resulting from the contribution of the ground states of the two Co spins with M$_z=\pm$3 or M$_z$=0. 

The presence of an exciton in the dot induces a spin-spin correlation with an enhancement of the occupation probability of the ferromagnetic ground states M$_z=\pm$3. The injection of spin polarized excitons permit an optical orientation of the two Co spins showing that some spin-flips among the exciton-2Co system are possible. Spin fluctuations under optical excitation in the tens of nanosecond range are revealed with auto-correlation measurements.

The four 2Co ground states can be addressed by resonant optical excitation of the exciton. The resonant excitation of the phonon side-band reveals absorption which strongly depends on a transverse magnetic field. These characteristic absorption feature arises from an interplay of the spin states mixing induced by the presence of an in-plane strain anisotropy and the transverse magnetic field. This tunable channel of pumping which strongly depends on the magnetic atoms fine structure $E/D_0$ could be used for a probing of the modulation of local strain in the vicinity of the magnetic atoms \cite{Tiwari2020JAP}.

\begin{acknowledgments}

We acknowledge the financial support from NationalScienceCenter Poland, project number 2021/41/B/ST3/04183.

\end{acknowledgments}

\end{document}